\documentclass{article}

\usepackage{arxiv}

\usepackage[T1]{fontenc}    
\usepackage{url}            
\usepackage[ toc, page, title, titletoc ]{appendix}
\usepackage{array}
\usepackage[ruled,noline]{algorithm2e}
\usepackage{amssymb}
\usepackage{amsmath}
\usepackage{amsfonts}
\usepackage{booktabs}
\usepackage[colorinlistoftodos]{todonotes}
\usepackage{hyperref}
\usepackage{lineno}
\usepackage{mathrsfs}
\usepackage{float}
\usepackage{url}            
\usepackage{makecell}
\usepackage{multirow}

\usepackage{etoolbox}
\providetoggle{fullBackground}

\usepackage{parskip}  

\DeclareFontFamily{U}{wncy}{}
\DeclareFontShape{U}{wncy}{m}{n}{<->wncyr10}{}
\DeclareSymbolFont{mcy}{U}{wncy}{m}{n}
\DeclareMathSymbol{\comb}{\mathord}{mcy}{"58} 

\usepackage{accents}
\newlength{\dhatheight}

\title{SPIRiT Regularization:\\ Parallel MRI with a Combination of\\ Sensitivity Encoding and Linear Predictability}

\author{
  Nicholas~Dwork\thanks{\url{www.nicholasdwork.com}, nicholas.dwork@cuanschutz.edu} \\
  Department of Biomedical Informatics \\
  University of Colorado Anschutz \\
  \And
  Alex~McManus \\
  Department of Applied Mathematics \\
  University of Colorado Boulder
  \And
  Stephen~Becker \\
  Department of Applied Mathematics \\
  University of Colorado Boulder
  \And
  Gennifer~T.~Smith \\
  Department of Engineering \\
  University of San Francisco
}

\begin{document}
\maketitle

\begin{abstract}
Accelerated Magnetic Resonance Imaging (MRI) permits high quality images from fewer samples that can be collected with a faster scan.  Two established methods for accelerating MRI include \emph{parallel imaging} and \emph{compressed sensing}.  Two types of parallel imaging include \emph{linear predictability}, which assumes that the Fourier samples are linearly related, and \emph{sensitivity encoding}, which incorporates \textit{a priori} knowledge of the sensitivity maps.  In this work, we combine compressed sensing with both types of parallel imaging using a novel regularization term: \emph{SPIRiT regularization}.  When combined, the reconstructed images are improved.  We demonstrate results on data of a brain, a knee, and an ankle.
\end{abstract}


\section{Introduction}

Magnetic Resonance Imaging (MRI) is a Fourier sensing modality, meaning that data are collected in the spatial frequency domain. However, compared to other imaging methods such as x-ray computed tomography (CT) or ultrasound, which offer comparable resolution, MRI is slow to collect data.  Whereas an unaccelerated 3D MRI scan for a volume $25\times 25\times 25$ cm$^3$ will require $>$5 minutes, the same volume can be imaged with CT in $<$10 seconds.  Its slow scan time makes the cost of each imaging study high and may necessitate the use of anesthesia for those patients who have difficulty remaining still (e.g., pediatric subjects or adult patients in pain).   One approach to accelerating MRI is to reduce the number of data samples collected.  The accompanying challenge is to reconstruct a high-quality image from those fewer data collected.  Established technologies for accelerating MRI in this way include partial Fourier sampling \cite{noll1991homodyne, mcgibney1993quantitative, bydder2005partial, mcmanus2024compressed}, parallel imaging \cite{deshmane2012parallel, pruessmann2001advances, pruessmann2006encoding}, compressed sensing \cite{donoho2005stable, lustig2007sparse, candes2008introduction}, and incorporating knowledge of the object's support \cite{plevritis1995spectral, samsonov2004pocsense, peng2011reference, dwork2024reducing}.  Partial Fourier sampling, parallel imaging, and compressed sensing are all approved by regulatory bodies and have been clinically deployed.  Parallel imaging, which takes advantage of multiple sensing coils with distinct sensitivity profiles, is now a mainstay clinical method for reducing scan time.

There are two main lines of research on parallel MRI: \textit{linear predictability} and \textit{sensitivity encoding}.  Parallel Imaging with Linear Predictability (PILP) assumes that each Fourier value is equal to a linear combination of nearby Fourier values \cite{sodickson1997simultaneous, griswold2002generalized, haldar2020linear, mcmanus2025dependence}.
Sensitivity encoding incorporates \textit{a priori} knowledge of the coils' sensitivity maps into the reconstruction \cite{pruessmann1999sense, pruessmann2001advances, pruessmann2001sensitivity, holme2019enlive}.  Often, with sensitivity encoding methods, the image is reconstructed by numerically solving an optimization problem \cite{fessler2020optimization}.  A common method for estimating sensitivity maps is ESPIRiT \cite{uecker2014espirit}, which is equivalent to PISCO \cite{lobos2023new}.

ESPIRiT is an extremely successful technique (available through the BART software package \cite{uecker2016bart}) which was described as a method ``where SENSE meets GRAPPA" \cite{uecker2014espirit}; SENSE was an early sensitivity encoding method \cite{pruessmann1999sense} and GRAPPA was an early PILP method \cite{griswold2002generalized}.  This description is fitting: ESPIRiT leverages the assumptions of PILP to estimate sensitivity maps, which are then used to reconstruct the image through a sensitivity encoding approach.
Though PILP and sensitivity encoding have been combined in this way, the combination is sequential: first, linear predictability is used to estimate the sensitivity maps and then, second, the image is reconstructed by solving an optimization problem.  Though the \textit{a priori} information included in both methods could conceivably be used to improve the quality of the reconstruction for a given number of samples, this is not currently the case.  Instead, PILP is used as a preprocessing step for a sensitivity encoding method.

In this manuscript, we present a single method that combines PILP with sensitivity encoding in such a way that they work synergistically to create a high-quality image.  We incorporate the assumption used by the SPIRiT PILP method as a regularization function into a sensitivity encoding method's optimization problem \cite{lustig2010spirit}.  By incorporating both methods into a single reconstruction process, the image quality for a given number of samples is improved.  Alternatively, the same quality could be achieved with fewer samples.

\section{Methods}

For this work, we are assuming that data will be collected on a Cartesian grid with two dimensions of phase encodes and one dimension of readout with all readout lines parallel to each other.  (This assumption is not necessary; future work will investigate non-Cartesian 3D data collection trajectories.)  The data is fully sampled along the readout direction and undersampled in the phase encode dimensions.  After data collection, an inverse Fourier transform is applied along each readout line to place the data into the hybrid $(k_x,k_y,z)$ domain \cite{beatty2007method}.  Once in the hybrid domain, each slice can be reconstructed independently; that is the approach we will take here.

SPIRiT \cite{lustig2010spirit} makes the fundamental assumption of PILP: any Fourier value equals a linear combination of nearby Fourier values with linear coefficients that are invariant across the frequency domain.  
In \cite{haldar2020linear}, Haldar et al. provide a necessary and sufficient condition for when PILP is accurate.  While valuable, it is difficult to relate this condition to the physical system.  In \cite{mcmanus2025dependence} it was shown that the PILP assumption is satisfied when there exists a linear combination of the coils' sensitivities such that the result is a complex exponential.  More specifically, consider the set of $c=1,2,\ldots,C$ coils.  PILP accurately interpolates the Fourier value for coil $\ell$ at spatial frequency $k=(k_x,k_y)$ whenever there exists a $d=(d_x,d_y)$ where the Fourier value at $k+d$ is known and there exist coefficients $n(d)\in\mathbb{C}^C$ such that
\begin{equation}
  \label{eq:smashAssumption}
  \sum_{c=1}^C n^{(c)}(d) \, S^{(c)}\left(x,y\right) \approx e^{i\left( d_x \, x + d_y \, y\right) } S^{(\ell)}\left(x,y\right),
\end{equation}
where $S^{(\ell)}$ is the sensitivity of the $\ell^{\text{th}}$ coil and $n^{(\ell)}(0)=0$.

SPIRiT uses a fully-sampled auto-calibration region (ACR) to estimate the optimal linear coefficients $\mathcal{W}$ for a set of directions that lie within a kernel.
Since the coefficients do not vary across the frequency domain, the process of interpolation can be expressed as sum of convolution operations:
\begin{equation}
  F S^{(\ell)} m = \sum_{c=1}^C \mathcal{W}^{(c)}_{(\ell)} \, \circledast \, F S^{(c)} m,
  \label{eq:spiritConditionIdeal}
\end{equation}
where $m\in\mathbb{C}^{U\times V}$ represents the image, $F$ represents the Discrete Fourier Transform ([DFT] - as defined in App. \ref{app:dft}), $\circledast$ represents circular convolution, and $\mathcal{W}^{(c)}_{(\ell)}$ is the kernel for interpolating the $\ell^{\text{th}}$ coil's Fourier values.  Since the sum of convolutions is a linear transformation, \eqref{eq:spiritConditionIdeal} can be expressed as
\begin{equation}
  \left( W^{(\ell)} - I \right) \, F \, S^{(\ell)} \, m = 0,
  \label{eq:spiritConditionIdealDiscrete}
\end{equation}
where $W^{(\ell)} \, k = \sum_{c=1}^C \mathcal{W}^{(c)}_{(\ell)} \circledast \, k$.  The values of $W^{(\ell)}$ are determined by solving the following least-squares optimization problem:
\begin{equation*}
  \underset{W^{(\ell)}}{\text{minimize}} \hspace{0.5em} \left\| W^{(\ell)} \ast k_{\text{ACR}} - k_{\text{interior}\left( \text{ACR} \right)} \right\|_2.
  \label{prob:findW}
\end{equation*}
Here, $\ast$ represents convolution and $k_{\text{ACR}}$ is a vector of the Fourier values of the spatial frequencies located within the ACR.  Problem \eqref{prob:findW} is a least-squares problem that can be solved with LSQR, which avoids explicitly forming the large matrix.  Note that here, we are using convolution instead of circular convolution and we are eliminating any values from the comparison where the kernel $W^{(\ell)}$ extends outside of the ACR.  See \cite{mcmanus2025dependence} for further details.

\subsection{Problem setup}

An image can be reconstructed with parallel imaging and compressed sensing by solving the Lagrangian form of a basis pursuit denoising optimization problem:
\begin{equation}
  \underset{m}{\text{minimize}} \hspace{0.5em} (1/2) \left\| \boldsymbol{M} \, \boldsymbol{F} \, \boldsymbol{S} \, m - \boldsymbol{b} \right\|_2^2 \hspace{0.25em}
  + \nu \| \Psi m \|_1.
  \label{prob:pics}
\end{equation}
Here, $\|\cdot\|_p$ is the $\ell_p$ norm, $\boldsymbol{S}=\left( S^{(1)}, S^{(2)}, \cdots, S^{(C)}\right)$ is a block-column matrix where each block element represents multiplication by a coil's sensitivity map, $\boldsymbol{F}=\text{diag}\left(F,F,\cdots,F\right)$ is a block-diagonal matrix that performs the DFT on each modulated image, $\boldsymbol{M}=\text{diag}(M,M,\cdots,M)$ is a block-diagonal matrix where $M$ is the data sampling mask that isolates the values for those frequencies that were collected, $\boldsymbol{b}=\left( b^{(1)}, b^{(2)}, \cdots b^{(c)} \right)$ is a complex block-column vector where $b^{(c)}$ is the vector of data collected by the $c^{\text{th}}$ coil, $\nu>0$ is the compressed sensing regularization parameter, and $\Psi$ is a sparsifying transformation.  The optimization variable $m$ is a complex vectorized image (i.e., the columns of the image are concatenated into a single vector).  For this work, we will use the Daubechies-4 discrete wavelet transformation as the sparsifying transformation \cite{majumdar2012choice}.

Equation \ref{eq:spiritConditionIdealDiscrete} is of the form that could be included as a regularization function in an optimization problem.  Note that \eqref{eq:spiritConditionIdealDiscrete} is true in the ideal case where \eqref{eq:smashAssumption} is perfectly satisfied; i.e., where the linear combination of sensitivity maps equals the $\ell^{\text{th}}$ sensitivity modulated by a perfect complex exponential.  However, it is unlikely that this is exactly the case.  Instead, the right-hand side of \eqref{eq:spiritConditionIdealDiscrete} will be nonzero but small.  With this understanding, parallel imaging with sensitivity encoding can be combined with PILP and compressed sensing as the following optimization problem:
\begin{equation}
  \underset{m}{\text{minimize}} \hspace{0.5em} (1/2) \left\| \boldsymbol{M} \, \boldsymbol{F} \, \boldsymbol{S} \, m - \boldsymbol{b} \right\|_2^2 \hspace{0.25em}
  + \nu \| \Psi m \|_1
  \hspace{0.25em} + \frac{\lambda_s}{2\kappa} \sum_{c=1}^C \left\| \left( W^{(c)} - I \right) \, F \, S^{(c)} \, m \right\|_{\gamma,2}^2.
  \label{prob:spiritReg}
\end{equation}
The new regularization term in \eqref{prob:spiritReg} is \emph{SPIRiT regularization}.
In problem \eqref{prob:spiritReg}, $\|\cdot\|_{\gamma,2}$ is the weighted $\ell_2$ norm, and $\lambda_s\geq 0$ is the spirit regularization parameter \footnote{Note that in this formulation, we are assuming that the samples collected are located on a Cartesian grid.  If this were not the case, then $M\,F$ would be replaced with a non-uniform DFT \cite{dutt1993fast, dutt1995fast, dwork2023optimization}}.  The weighted $\ell_2$ norm is defined as follows:
\begin{equation*}
  \| x \|_{\gamma,2} = \sqrt{ \gamma_1 |x_1|^2 + \gamma_2 |x_2|^2 + \cdots + \gamma_N |x_N|^2 }.
\end{equation*}
The value of $\kappa$ is used to balance the terms of the objective function and bring the appropriate value for $\lambda_s$ closer to $1$; it is set to the square root of the ratio of matrix induced norms:
\begin{equation*}
  \kappa = \sqrt{ \frac{ 
  \left\| \text{diag}\left( \left( W^{(1)} - I \right) \, F \, S^{(1)}, 
                            \left( W^{(2)} - I \right) \, F \, S^{(2)}, \ldots, 
                            \left( W^{(C)} - I \right) \, F \, S^{(C)} \right) \right\| }{
          \left\|\boldsymbol{M} \, \boldsymbol{F} \, \boldsymbol{S} \right\| } }.
\end{equation*}

In \eqref{prob:spiritReg}, without using a weighted norm (i.e., $\gamma=(1,1,\ldots,1)$), the optimization algorithm could make the errors for large frequencies as large as the errors for small frequencies to yield an optimal solution.  As a percentage of the true Fourier magnitudes, these errors would be a much larger percentage of the Fourier values in the high frequencies than in the low frequencies.  To mitigate this outcome, we will alter the weights of the weighted norm.

The magnitude spectrum of natural images (without noise) for frequencies other than $0$ tends to follow a power law \cite{bracewell1995two, ruderman1997origins, van1996modelling}.  (The power law would predict infinite power for the $0$ frequency, which is never valid for natural images.)  
We will weight the norm with the inverse of a fit based on the power law.  To accommodate the presence of noise, we will fit small and large frequencies to power laws independently and then take the maximum of the fit, as follows:
\begin{equation*}
  P(k) = \text{max}\left( m_L \cdot |k|^{-p_L}, m_H \cdot |k|^{-p_H} \right).
\end{equation*}
To identify the low and high scaling coefficients $m_L$ and $m_H$, and the corresponding exponents $p_L$ and $p_H$, we propose use use the Levenberg-Marquardt optimization algorithm \cite{boyd2018introduction} to find the values that minimize the $\ell_2$ norm of the difference between $P(k)$ and $|b(k)|$ for all collected spatial frequencies.  To identify the value of the fit at $k=0$, we fit a line to a small number of points surrounding the $0$ frequency.  Once the fit is complete, the values of $\gamma$ are set to the inverse of the fitted values:
\begin{equation}
  \gamma_i = P(k_i)^{-1} = \text{min}\left( \frac{|k_i|^{p_L} }{ m_L }, \frac{ |k_i|^{p_H} }{ m_H } \right).
  \label{eq:spiritRegNormWeights}
\end{equation}
This weighting is higher for the high spatial frequencies and lower for the low spatial frequencies.  It ensures that the errors in the high frequencies are much smaller than those in the low frequencies, commensurate with the magnitude of the Fourier values.  Reconstructing an MR image by solving problem \eqref{prob:spiritReg} with weights determined according to \eqref{eq:spiritRegNormWeights} is parallel imaging with compressed sensing and SPIRiT regularization (PICS+SR).

\subsection{Solving the optimization problems}

The optimization problem of \eqref{prob:spiritReg} can be solved with the Fast Iterative Shrinkage Thresholding Algorithm (FISTA) \cite{beck2009fast}.  Specifically, we will use FISTA with line search as detailed in \cite{o2018fraction}.  FISTA solves problems of the form
\begin{equation}
  \text{minimize} \hspace{0.5em} G(y) + H(y),
  \label{eq:fistaForm}
\end{equation}
where $G$ is convex and differentiable, and $H$ is closed, convex, and proper (CCP) with a computationally feasible proximal operator, defined as follows:
\begin{equation*}
  \text{prox}_{tG}(x) = \underset{v}{\text{argmin}} \hspace{0.25em} G(v) + \frac{1}{2t}\left\|x-v\right\|_2^2.
\end{equation*}

As an example, consider the PICS problem of \eqref{prob:pics}.  Let $G(m) = (1/2) \| \boldsymbol{M}\boldsymbol{F}\boldsymbol{S} m - \boldsymbol{b} \|_2^2$ and let $H(m)=\nu\|\Psi m\|_1$.  Then $G$ is differentiable, $H$ has a simple proximal operator when $\Psi$ is tight frame, and problem \eqref{eq:fistaForm} is the PICS problem, which can now be solved with FISTA.  The proximal operator of $H$ is soft-thresholding applied to the wavelet coefficients of $m$ \cite{combettes2005signal}.  We will let $\Psi$ be the discrete Daubechies-4 wavelet transform, which is orthogonal (and therefore tight frame), so $H$ has a simple proximal operator.

If we instead define $G(m) = (1/2) \| \boldsymbol{M}\boldsymbol{F}\boldsymbol{S} m - \boldsymbol{b} \|_2^2 + (\lambda_s/2) \sum_{c=1}^C \left\| \left( W^{(c)} - I \right) \, F \, S^{(c)} \, m \right\|_{\gamma,2}^2$ with the same definition of $H$, then $G$ is differentiable, $H$ still has a simple proximal operator, and \eqref{eq:fistaForm} becomes the PICS+SR problem of \eqref{prob:spiritReg}, which can be solved with FISTA.

\section{Experiments}

\begin{figure}[ht]
  \centering
  \includegraphics[width=0.75\linewidth]{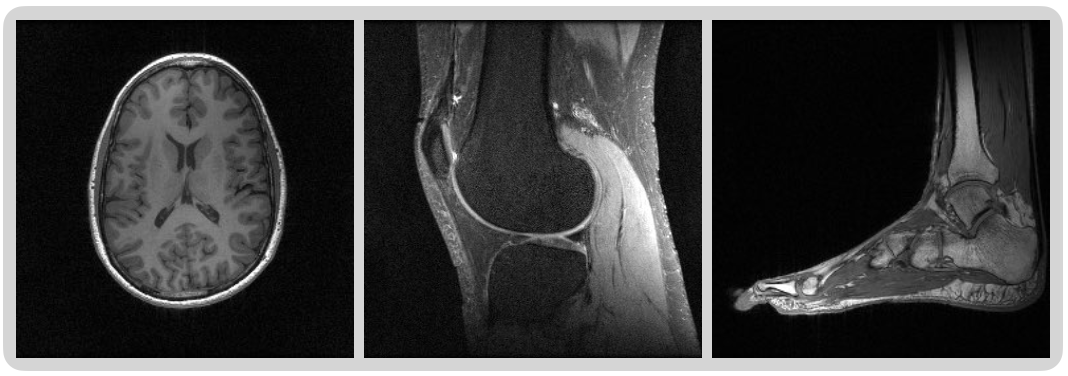}
  \caption{Fully-sampled reconstructions of the brain, knee, and ankle images analyzed in this work.}
  \label{fig:trueImgs}
\end{figure}

We demonstrate the effectiveness of the proposed algorithm on MRI data of a knee (made public through mridata.org \cite{ong2018mridata}), and a brain and an ankle (made public with \cite{dwork2024accelerated}); all images were size $256\times 256$.  The sensitivity maps were identified with PISCO \cite{lobos2023new}.  Figure \ref{fig:trueImgs} shows the fully-sampled reconstructions.

\begin{figure}[ht]
  \centering
  \includegraphics[width=0.9\linewidth]{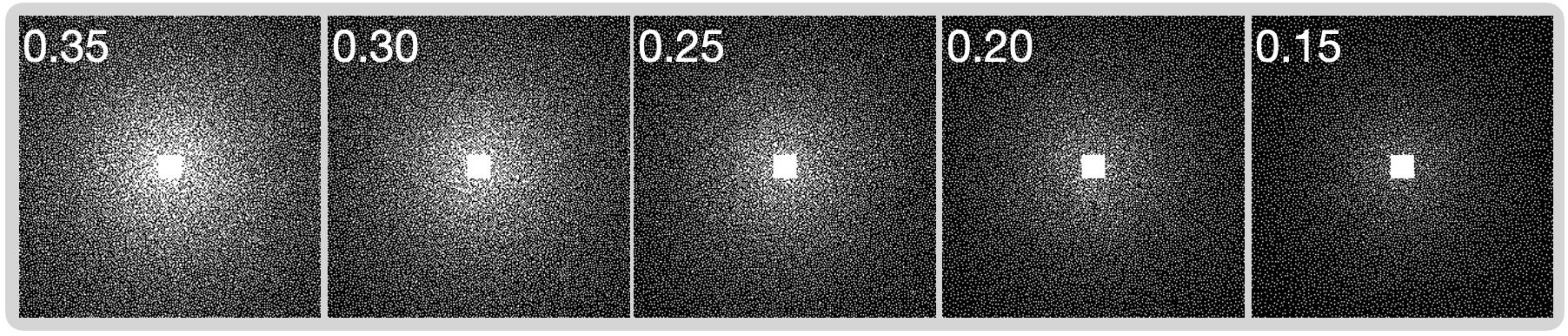}
  \caption{The sampling patterns used in this work ranging from sample fractions of 0.15 to 0.35 where a white spot indicates a sample that was collected and a black spot indicates a sample that was not.  The white numbers indicate the fraction of all samples that were collected.  The white square in the middle is the auto-calibration region (ACR).}
  \label{fig:samplingPatterns}
\end{figure}

All data were collected with 3DFT trajectories; i.e., the samples were located on a Cartesian grid and there were two dimensions of phase encoding with one dimension of readout.
To assess the quality that could be attained with accelerated MRI, the data was retrospectively downsampled according to a variable density Poisson disc sampling pattern that included a fully-sampled region, generated according to \cite{dwork2021fast}.  The size of the fully-sampled region centered on the $0$ frequency -- the autocalibration region (ACR) -- was determined according to the two-level sampling pattern of \cite{adcock2017breaking} and was used in \cite{dwork2021utilizing, dwork2022utilizing, dwork2024accelerated}; with the four scales that we used for the wavelet transformation, the size of the finest scale is $16\times 16$.  The sampling patterns for the different sample fractions studied is shown in Fig. \ref{fig:samplingPatterns}.

To quantify image quality, we will compare the reconstructed image to the fully-sampled reconstruction: the result of solving problem \eqref{prob:pics} with $\nu=0$ without any downsampling.  We will compute the Structural Similarity Index Measure (SSIM) \cite{wang2004image} between the magnitude of the reconstructed image and the fully-sampled reconstruction, and we will compute the peak signal-to-noise ratio (PSNR) between the two complex images \cite{ong2018low}.

\section{Results}

Images were reconstructed for all combinations of values of $\nu~\in~\{0.01, 0.1, 1, 10\}$ and $\gamma~\in~\{0.1, 0.5, 1, 2, 4, 5, 10\}$.  Results for the reconstructions that yielded the highest value of SSIM are reported here.  Quality metric values for all three datacases -- the brain, knee, and ankle -- are shown in Table \ref{tbl:quality_metricsSR}.  For all results, the optimization algorithm was run for 1000 iterations; the difference in objective value between 500 and 1000 iterations was negligible (data not shown), indicating convergence. Incorporating our novel SPIRiT regularization consistently improved both SSIM and PSNR.

\begin{table}[h]
\centering
\caption{Quality Metric Values}
\label{tbl:quality_metricsSR} 
\begin{tabular}{>{\centering\arraybackslash}m{1.5cm} c l *{2}{c}}
    \toprule
    \textbf{Dataset} & \textbf{Metric} & \textbf{Sample Fraction} & \textbf{PICS} & \textbf{PICS+SR} \\
    \midrule
    \multirow{8}{*}{\rotatebox{90}{Brain}} & \multirow{4}{*}{SSIM} & 0.35 & 0.966 & \underline{0.967} \\
    & & 0.3 & 0.962 & \underline{0.964} \\
    & & 0.25 & \underline{0.959} & \underline{0.959} \\
    & & 0.2 & 0.952 & \underline{0.952} \\
    \cmidrule(lr){2-5}
    & \multirow{4}{*}{PSNR (dB)} & 0.35 & 40.15 & \underline{40.24} \\
    & & 0.3 & 39.60 & \underline{39.69} \\
    & & 0.25 & 39.06 & \underline{39.15} \\
    & & 0.2 & 38.38 & \underline{38.47} \\
    \midrule
    \multirow{8}{*}{\rotatebox{90}{Knee}} & \multirow{4}{*}{SSIM} & 0.35 & 0.800 & \underline{0.814} \\
    & & 0.3 & 0.779 & \underline{0.796} \\
    & & 0.25 & 0.750 & \underline{0.772} \\
    & & 0.2 & 0.721 & \underline{0.747} \\
    \cmidrule(lr){2-5}
    & \multirow{4}{*}{PSNR (dB)} & 0.35 & 29.67 & \underline{30.42} \\
    & & 0.3 & 29.12 & \underline{29.93} \\
    & & 0.25 & 28.49 & \underline{29.30} \\
    & & 0.2 & 27.89 & \underline{28.67} \\
    \midrule
    \multirow{8}{*}{\rotatebox{90}{Ankle}} & \multirow{4}{*}{SSIM} & 0.35 & 0.974 & \underline{0.975} \\
    & & 0.3 & 0.970 & \underline{0.971} \\
    & & 0.25 & 0.966 & \underline{0.967} \\
    & & 0.2 & 0.960 & \underline{0.961} \\
    \cmidrule(lr){2-5}
    & \multirow{4}{*}{PSNR (dB)} & 0.35 & 40.74 & \underline{40.83} \\
    & & 0.3 & 39.94 & \underline{40.03} \\
    & & 0.25 & 39.06 & \underline{39.16} \\
    & & 0.2 & 37.95 & \underline{38.06} \\
    \bottomrule
\end{tabular}
\vspace*{6pt} \\ 
\parbox{0.75\linewidth}{Quality metric values for reconstructions using parallel imaging and compressed sensing (PICS) and PICS with SPIRiT regularization (PICS+SR).  The best reconstruction with respect to each metric is underlined.}
\end{table}

\begin{figure}[ht]
  \centering
  \includegraphics[width=0.8\linewidth]{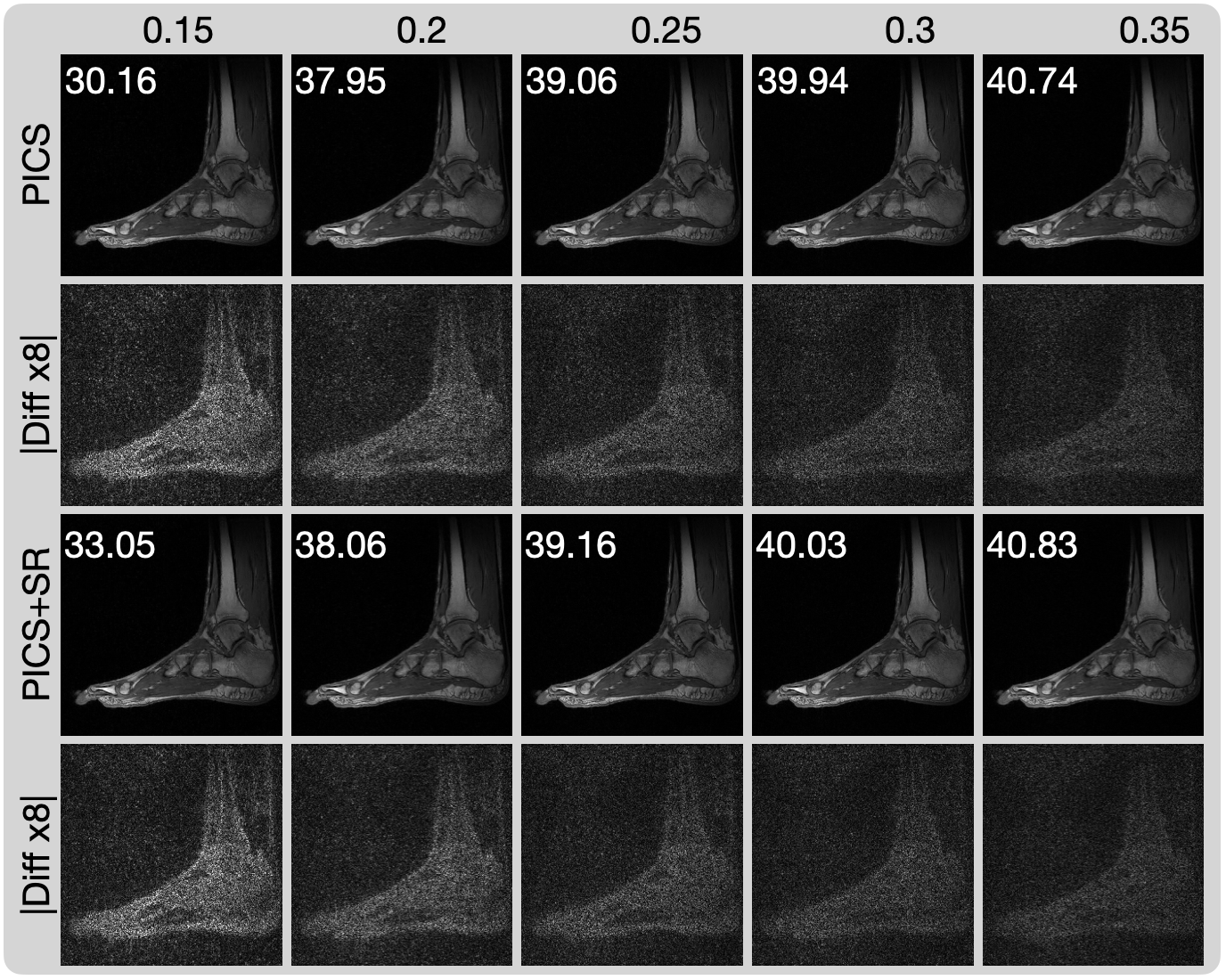}
  \caption{Reconstructions of the ankle for several different sampling fractions.  The first and third rows show the reconstructions with parallel imaging and compressed sensing (PICS), and PICS with SPIRiT regularization, respectively.  The second and fourth rows show the magnitude of the difference images multiplied by 8.  The PSNR in decibels for each reconstruction is also presented in white.}
  \label{fig:ankleReconsAblation}
\end{figure}

Figure \ref{fig:ankleReconsAblation} shows reconstructions of the ankle with and without SPIRiT regularization (SR) for several different sampling fractions.  As expected, the image quality of PICS and PICS+SR improves with sample fraction.  For all sample fraction, the inclusion of SPIRiT regularization improves the image quality.   The improvement in image quality when using SPIRiT regularization is subtle and presents as a general reduction in the noise across the image (rather than an improvement in any localized structural features).  The improvement with SPIRiT regularization is more significant at lower sample fractions.

\begin{figure}[ht]
  \centering
  \includegraphics[width=0.5\linewidth]{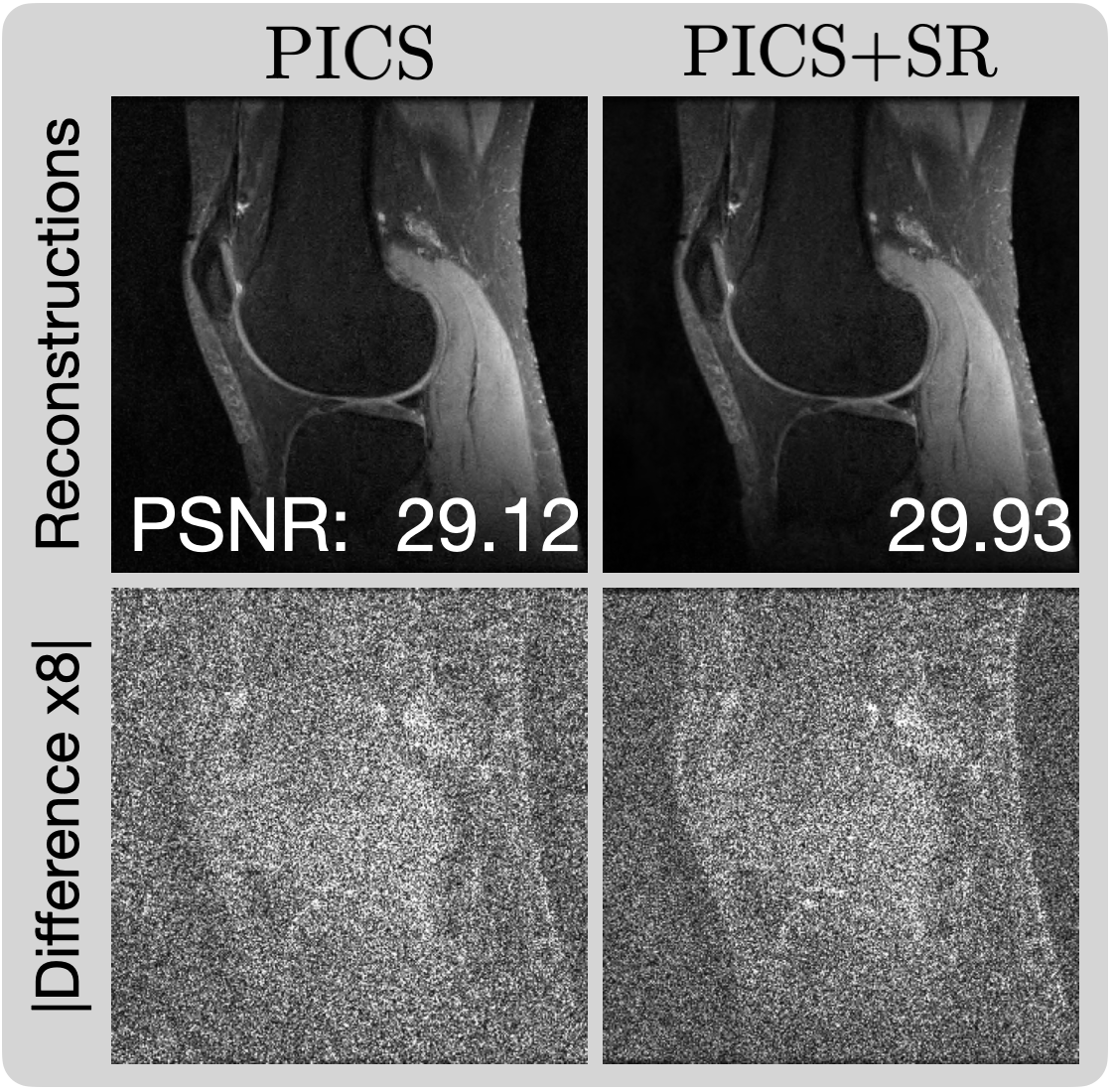}
  \caption{Reconstructions of the knee with a 0.3 sample fraction using parallel imaging and compressed sensing (PICS) and PICS with SPIRiT regularization (SR).  The second row shows the magnitudes of the differences between the under-sampled reconstructions and the fully-sampled reconstructions multiplied by 8.  The PSNR in decibels for each reconstruction is also presented in white.}
  \label{fig:kneeReconsSR0.3}
\end{figure}

Figure \ref{fig:kneeReconsSR0.3} shows the reconstructions of the knee with a 0.30 sample fraction with the magnitudes of the differences multiplied by 8.  The difference images show the improvement gained by including SPIRiT regularization in the reconstruction.

\begin{figure}[ht]
  \centering
  \includegraphics[width=0.5\linewidth]{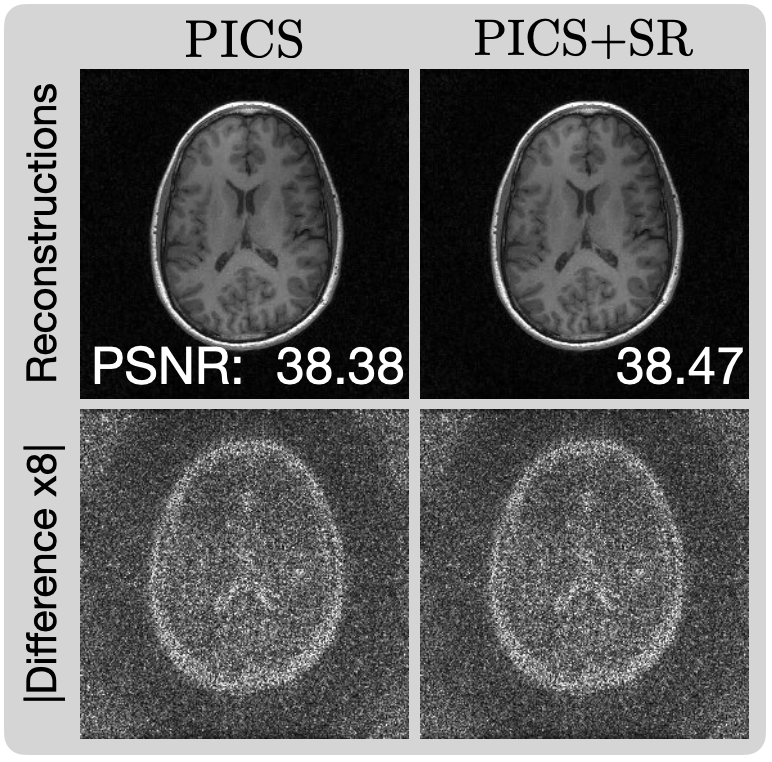}
  \caption{Reconstructions of the brain with a 0.2 sample fraction using parallel imaging and compressed sensing (PICS), PICS with SPIRiT regularization (SR).  The second row shows the magnitudes of the differences between the under-sampled reconstructions and the fully-sampled reconstructions multiplied by 8.  The PSNR in decibels for each reconstruction is also presented in white.}
  \label{fig:brainReconsSR0.2}
\end{figure}

Figure \ref{fig:brainReconsSR0.2} shows the reconstructions of the brain with a 0.20 sample fraction along with the magnitudes of the differences multiplied by 8.  The improvement in the brain with SPIRiT regularization is more subtle than that of the knee, as can be seen in the difference images.

\section{Conclusion}

When combined with PICS, SPIRiT regularization improves image quality over the use of PICS alone.  The improvement is more significant at lower sample fractions and less so at higher sample fractions.  Importantly, the only change comes in the form of the reconstruction algorithm; the process of data collection remains the same.  Thus, this improvement comes without any additional scan time or modification to the clinical workflow.

In all cases, for both PICS and PICS+SR, the $\lambda$ and $\gamma$ values that yielded the optimal PSNR values were $0.1$ and $0.5$.  This suggests that we could use the same values for new datasets and expect a high quality reconstruction.  However, a study of much more data would be required to make such a conclusion.

For this work, we investigated an additional modification where knowledge of the support was included in the reconstruction process; our approach is detailed in App. \ref{app:support}.  The support is the set of pixels that image anatomy.
While inclusion of the support did not improve image quality significantly in this work, in past work, it was shown that the sampling pattern could be reduced using the support alone \cite{dwork2024reducing}.  Thus, it may be more beneficial to incorporate knowledge of the support with something other than a variable density Poisson disc sampling pattern; perhaps a variable density subsampling of the pattern presented in \cite{dwork2024reducing} would be more appropriate.  Separately, the method of \cite{lobos2023new} identifies a conservative superset of the support.  In particular, for the knee, only a small portion of the region outside of the knee is excluded from the support (see Fig. \ref{fig:supports} in App. \ref{app:support}).  A different method that identifies more of the region outside of the support (i.e., outside of the anatomy) may improve the effect of including the support in the reconstruction method.  We will explore these ideas in future work.

In this work, we fixed the number of iterations of the optimization algorithms to 1000.  Rather than having a fixed number of iterations for each optimization algorithm, it would be beneficial to include a dynamic stopping criteria that would automatically stop the optimization algorithms when close to convergence.  This would reduce the time to reconstruct the image without any reduction in image quality.  However, this would likely include an additional parameter for the threshold at which to stop the iterations.  Again, we will investigate this approach in future work.

Other future efforts will alter the optimization problem presented in this manuscript to incorporate partial Fourier sampling \cite{bydder2005partial}, similar to \cite{mcmanus2024compressed}, and structured sparsity \cite{dwork2021utilizing, dwork2022utilizing, dwork2024accelerated}.

\section*{Acknowledgments}
This work was supported in part by a Rifkin and Bennis Cancer Imaging AI Early-stage Project Award.

\appendices

\section{Discrete Fourier Transform}
\label{app:dft}
The definition of the two-dimensional discrete Fourier transform $F$ used in this manuscript is
\begin{equation*}
    F\,x = \sum_{u=0}^{U-1}\sum_{v=0}^{V-1} x_{u,v} \exp\left( -i\,2\pi\,\left(\frac{k_u\,u}{U} + \frac{k_v\,v}{V} \right)\right).
\end{equation*}
The inverse Discrete Fourier transform is
\begin{equation*}
  F^{-1}\,y = \frac{1}{UV} \sum_{k_u=0}^{U-1} \sum_{k_v=0}^{V-1} \hat{x}_{u,v} \exp\left( i\,2\pi\,\left(\frac{k_u\,u}{U} + \frac{k_v\,v}{V} \right) \right),
\end{equation*}
where $\hat{x}=F\,x$.

\section{Inclusion of the Support}
\label{app:support}

The process of clinical MRI consists of 1) reconstructing a low-resolution \textit{localizer} image to ensure the patient is properly positioned in the bore and from which the technologist selects the field-of-view ([FOV] - the volume of space to image with high fidelity), and 2) reconstruct a high-quality image for the volume within the FOV from a second set of collected data.  Conventionally, the FOV selected by the technologist has been a rectangle.  However, the system could be altered so that the technologist could indicate the support of the object to be imaged (i.e., a non-rectangular region that contains the object).  The FOV would then be the smallest rectangle that contains this support.  Alternately, an automatic system -- such as PISCO \cite{lobos2023new} -- could identify the non-rectangular support from the data used to reconstruct the localizer image.  The support, which encloses a smaller volume than the FOV, could require less data to reconstruct a high-quality image \cite{dwork2024reducing}.

Let the support of the image be the set of those locations in space where the true image (without noise) is non-zero.  Let $\Omega$ be a superset of the support.  In past works, this superset has been used to improve image quality for a given number of samples \cite{samsonov2004pocsense}.  Let $T_{\bar{\Omega}}$ be the linear transformation that extracts those pixels outside of the support $\Omega$ and arranges them into a vector.  (Here, $\bar{\Omega}$ is the complement of the set of pixels within the support.)  Then, problem \eqref{prob:spiritReg} is modified to include the support as follows.

\begin{equation}
  \begin{aligned}
    \underset{m}{\text{minimize}} &\hspace{0.5em} (1/2) \left\| \boldsymbol{M} \, \boldsymbol{F} \, \boldsymbol{S} \, m - \boldsymbol{b} \right\|_2^2 \hspace{0.5em}
    + \nu \| \Psi m \|_1
    + (\lambda_s/2) \sum_{c=1}^C \left\| \left( W^{(c)} - I \right) \, F \, S^{(c)} \, m \right\|_{\gamma,2}^2 \\
    \text{subject to} &\hspace{0.5em} \frac{ \| T_{\bar{\Omega}} \, m \|_2^2 }{ |\bar{\Omega}| } \leq \sigma^2.
    \label{prob:spiritRegWithSupport}
  \end{aligned}
\end{equation}
Here, $|\bar{\Omega}|$ is the number of elements in $\bar{\Omega}$, and $\sigma^2$ is a bound on the average energy for those pixels outside of the support.  Unlike the regularization parameters $\nu$ and $\lambda_s$, $\sigma^2$ can be estimated from data by computing the sample variance of an image reconstructed from data captured during an acquisition without any excitation.

The optimization problem of \eqref{prob:spiritRegWithSupport} can be solved with the primal-dual hybrid gradient (PDHG) method \cite{chambolle2011first}.  Specifically, we will use PDHG with line search, as detailed in \cite{malitsky2018first}.  PDHG solves problems of the form
\begin{equation}
  \text{minimize} \hspace{0.5em} V(y) + W(A\,y),
  \label{eq:pdhgForm}
\end{equation}
where $V$ and $W$ are both CCP with computationally feasible proximal operators and $A$ is a linear transformation.

To put problem \eqref{prob:spiritRegWithSupport} is put in the form of \eqref{eq:pdhgForm}, first note that \eqref{prob:spiritRegWithSupport} is equivalent to
\begin{equation}
  \begin{aligned}
    \underset{m}{\text{minimize}} &\hspace{0.5em} \frac12\,\left\| \boldsymbol{M} \, \boldsymbol{F} \, \boldsymbol{S} \, m - \boldsymbol{b} \right\|_2^2 
      + (\lambda_s/2) \sum_{c=1}^C \left\| \left( W^{(c)} - I \right) \, F \, S^{(c)} \, m \right\|_{\gamma,2}^2 \\
      &+ \nu \|\Psi m\|_1
      + \mathbb{I}_{B\left[0,\sigma \sqrt{|\Omega|} \right]}( P_{\bar{\Omega}} \, m ),
  \end{aligned}
  \label{prob:picsSpiritRegEquiv}
\end{equation}
where $B\left[0,\sqrt{\nu\,|\Omega|} \right]$ is the closed ball centered on $0$ with radius $\sqrt{\nu\,|\Omega|}$.  The function $\mathbb{I}_Q(\cdot)$ is the indicator function for the set $Q$: equal to $0$ when its argument is within the set and otherwise equal to infinity.

Let $V(m)=\|\Psi m\|_1$, 
\begin{equation*}
  A=\begin{bmatrix}
    \boldsymbol{M}\boldsymbol{F}\boldsymbol{S} \\
    \left( W^{(1)} - I \right) \, F \, S^{(1)}) \\
    \left( W^{(2)} - I \right) \, F \, S^{(2)}) \\
    \vdots \\
    \left( W^{(C)} - I \right) \, F \, S^{(C)}) \\
    I
    \end{bmatrix}
\end{equation*}
and
\begin{equation*}
  W(y^{(1)},y^{(2)},\ldots,y^{(C+1)},y^{(C+2)}) = \frac{1}{2}\|y^{(1)}-\boldsymbol{b}\|_2^2
    + \frac{\lambda_s}{2} \sum_{c=1}^C \left\| y^{(c+1)} \right\|_{\gamma,2}^2 + \mathbb{I}_{B\left[0,\sqrt{\nu\,|\Omega|} \right]}( P_{\bar{\Omega}} \, m ).
\end{equation*}
It has already been established that $V$ is CCP with a simple proximal operator.  Similarly, $W$ is CCP.  Since each term in the sum of $W$ has a simple proximal operator, and since each term operates on different components of $y$, $W$ has a simple proximal operator by the separate sum property of proximal operators.  Therefore, problem \eqref{eq:pdhgForm} is equivalent to \eqref{prob:picsSpiritRegEquiv} which can be solved with PDHG.

\begin{figure}[ht]
  \centering
  \includegraphics[width=0.6\linewidth]{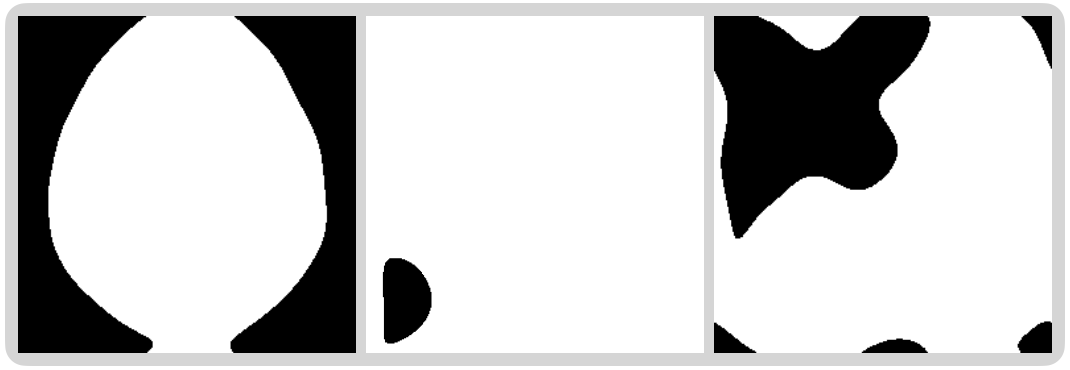}
  \caption{ The support regions identified with PISCO for the brain (left), the knee (center), and the ankle(right). }
  \label{fig:supports}
\end{figure}

For this work, since data collected without excitation was unavailable, $\sigma^2$ was estimated using the sample variance of a region of the fully-sampled reconstruction that was outside of the anatomy.  For this work, the support regions were identified with PISCO \cite{lobos2023new}; they are shown in Fig. \ref{fig:supports}.

\begin{table}[h]
\centering
\caption{Quality Metric Values}
\label{tbl:quality_metrics} 
\begin{tabular}{>{\centering\arraybackslash}m{1.5cm} c l *{4}{c}}
    \toprule
    \textbf{Dataset} & \textbf{Sample Fraction} & \textbf{Metric} & \textbf{PICS} & \textbf{PICS+SR} & \textbf{PICS+$\Omega$} & \textbf{PICS+SR+$\Omega$} \\
    \midrule
    \multirow{8}{*}{\rotatebox{90}{Brain}} & \multirow{4}{*}{SSIM} & 0.35 & 0.966 & 0.967 & \underline{0.967} & \underline{0.967} \\
    & & 0.3 & 0.962 & \underline{0.964} & 0.963 & \underline{0.964} \\
    & & 0.25 & 0.959 & \underline{0.959} & 0.959 & 0.959 \\
    & & 0.2 & 0.952 & \underline{0.952} & 0.952 & 0.952 \\
    \cmidrule(lr){2-7}
    & \multirow{4}{*}{PSNR (dB)} & 0.35 & 40.15 & \underline{40.24} & 40.15 & \underline{40.24} \\
    & & 0.3 & 39.60 & \underline{39.69} & 30.60 & \underline{39.69} \\
    & & 0.25 & 39.06 & \underline{39.15} & 39.06 & \underline{39.15} \\
    & & 0.2 & 38.38 & \underline{38.47} & 38.38 & \underline{38.47} \\
    \midrule
    \multirow{8}{*}{\rotatebox{90}{Knee}} & \multirow{4}{*}{SSIM} & 0.35 & 0.800 & \underline{0.814} & 0.800 & \underline{0.814} \\
    & & 0.3 & 0.779 & 0.796 & 0.779 & \underline{0.7964} \\
    & & 0.25 & 0.750 & 0.772 & 0.750 & \underline{0.772} \\
    & & 0.2 & 0.721 & \underline{0.747} & 0.721 & 0.746 \\
    \cmidrule(lr){2-7}
    & \multirow{4}{*}{PSNR (dB)} & 0.35 & 29.66 & \underline{30.42} & 29.66 & \underline{30.42} \\
    & & 0.3 & 29.12 & \underline{29.93} & 29.12 & 29.92 \\
    & & 0.25 & 28.49 & \underline{29.30} & 28.49 & \underline{29.30} \\
    & & 0.2 & 27.89 & \underline{28.67} & 27.89 & 28.66 \\
    \midrule
    \multirow{8}{*}{\rotatebox{90}{Ankle}} & \multirow{4}{*}{SSIM} & 0.35 & 0.974 & \underline{0.975} & 0.974 & \underline{0.975} \\
    & & 0.3 & 0.970 & \underline{0.971} & 0.970 & \underline{0.971} \\
    & & 0.25 & 0.966 & \underline{0.967} & 0.966 & 0.967 \\
    & & 0.2 & 0.960 & \underline{0.961} & 0.960 & \underline{0.961} \\
    \cmidrule(lr){2-7}
    & \multirow{4}{*}{PSNR (dB)} & 0.35 & 40.74 & \underline{40.83} & 40.74 & \underline{40.83} \\
    & & 0.3 & 39.94 & \underline{40.03} & 39.94 & \underline{40.03} \\
    & & 0.25 & 39.06 & \underline{39.16} & 39.06 & \underline{39.16} \\
    & & 0.2 & 37.95 & \underline{38.06} & 37.95 & \underline{38.06} \\
    \bottomrule
\end{tabular}
\vspace*{6pt} \\ 
\parbox{0.75\linewidth}{Quality metric values for reconstructions using parallel imaging and compressed sensing (PICS), PICS with SPIRiT regularization (SR), PICS with the support ($\Omega$), and PICS with SPIRiT regularization as well as the support.  The best reconstruction with respect to each metric is underlined.}
\end{table}

In all cases, the difference in metric value when including the support is extremely small (and perhaps negligible).  With respect to PSNR, the inclusion of the support improves the outcome in the majority of reconstructions.  With respect to SSIM, though, the inclusion of the support yields the best metric value in only half of the reconstructions.

\begin{figure}[ht]
  \centering
  \includegraphics[width=0.8\linewidth]{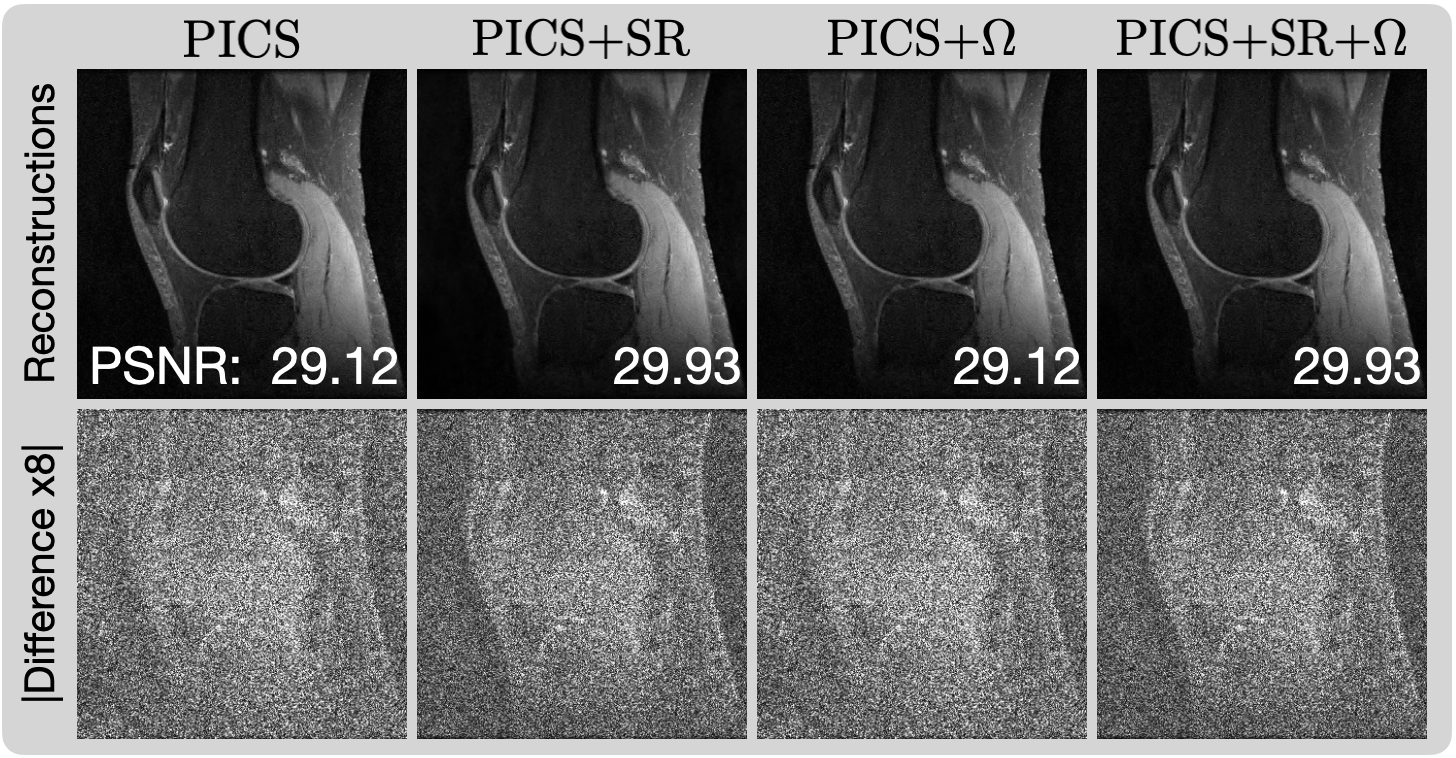}
  \caption{Reconstructions of the knee with a 0.3 sample fraction using parallel imaging and compressed sensing (PICS), PICS with SPIRiT regularization (SR), PICS with the support ($\Omega$), and PICS with SPIRiT regularization as well as the support.  The second row shows the magnitudes of the differences between the under-sampled reconstructions and the fully-sampled reconstructions multiplied by 8.  The PSNR for each reconstruction is also presented.}
  \label{fig:kneeRecons0.3}
\end{figure}

\begin{figure}[ht]
  \centering
  \includegraphics[width=0.8\linewidth]{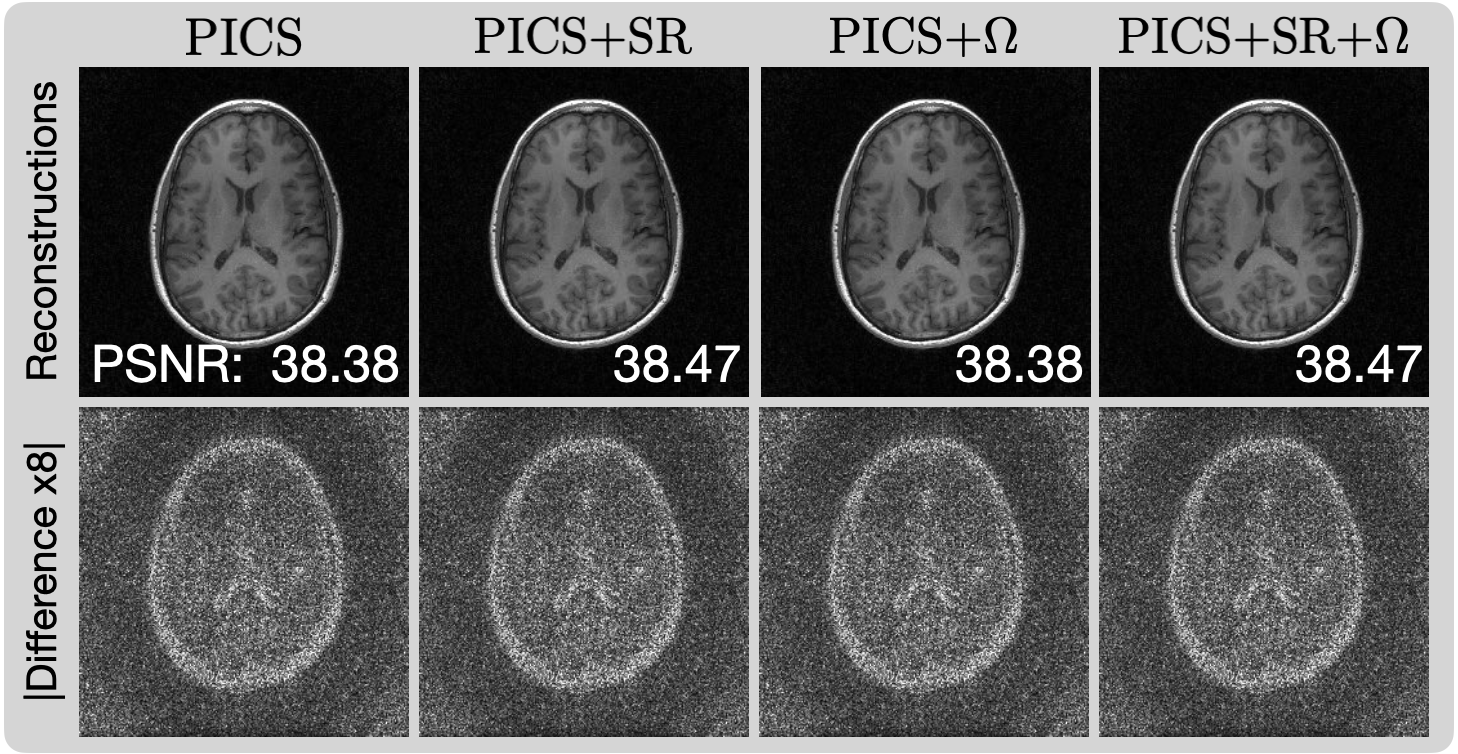}
  \caption{Reconstructions of the brain with a 0.2 sample fraction using parallel imaging and compressed sensing (PICS), PICS with SPIRiT regularization (SR), PICS with the support ($\Omega$), and PICS with SPIRiT regularization as well as the support.  The second row shows the magnitudes of the differences between the under-sampled reconstructions and the fully-sampled reconstructions multiplied by 8.  The PSNR for each reconstruction is also presented in decibels.}
  \label{fig:brainRecons0.2}
\end{figure}

\begin{figure}[ht]
  \centering
  \includegraphics[width=0.8\linewidth]{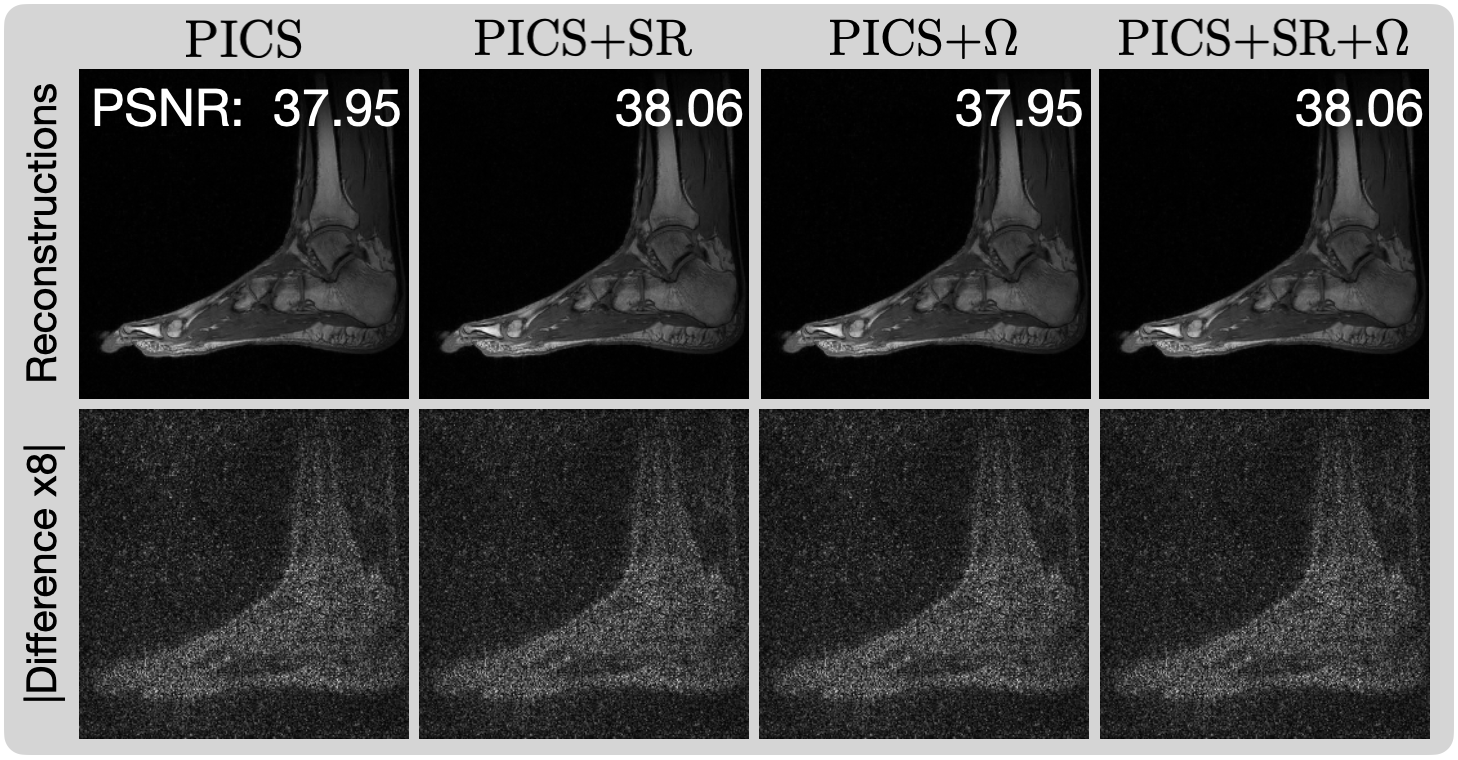}
  \caption{Reconstructions of the ankle with a 0.2 sample fraction using parallel imaging and compressed sensing (PICS), PICS with SPIRiT regularization (SR), PICS with the support ($\Omega$), and PICS with SPIRiT regularization as well as the support.  The second row shows the magnitudes of the differences between the under-sampled reconstructions and the fully-sampled reconstructions multiplied by 8.  The PSNR for each reconstruction is also presented in decibels.}
  \label{fig:ankleRecons0.2}
\end{figure}

Figures \ref{fig:kneeRecons0.3} shows reconstructions of the knee with a sample fraction of 0.3, and figures \ref{fig:brainRecons0.2}, and \ref{fig:ankleRecons0.2} show the reconstructions of the brain and ankle with a 0.20 sample fraction for the four different reconstruction methods along with the magnitudes of the differences from the fully-sampled images multiplied by 8.  The difference images show that the improvement in image quality when including the support is negligible.

Note that problem \eqref{prob:spiritRegWithSupport} contains a differentiable term, making it amenable to being solved with the Condat-Vu algorithm \cite{condat2013primal} or with PD3O \cite{yan2016primal}.  Should a sampling pattern or support region selection system be discovered that improves image quality when incorporating the support, these alternatives to PDHG may converge to a high-quality solution faster.


\begin{thebibliography}{10}

\bibitem{noll1991homodyne}
Douglas~C Noll, Dwight~G Nishimura, and Albert Macovski.
\newblock Homodyne detection in magnetic resonance imaging.
\newblock {\em IEEE transactions on medical imaging}, 10(2):154--163, 1991.

\bibitem{mcgibney1993quantitative}
G~McGibney, MR~Smith, ST~Nichols, and A~Crawley.
\newblock Quantitative evaluation of several partial fourier reconstruction
  algorithms used in {MRI}.
\newblock {\em Magnetic resonance in medicine}, 30(1):51--59, 1993.

\bibitem{bydder2005partial}
Mark Bydder and Matthew~D Robson.
\newblock Partial fourier partially parallel imaging.
\newblock {\em Magnetic Resonance in Medicine}, 53(6):1393--1401, 2005.

\bibitem{mcmanus2024compressed}
Alex McManus, Stephen~R Becker, Daniel O'Connor, and Nicholas Dwork.
\newblock Compressed sensing with homodyne detection.
\newblock In {\em 2024 IEEE Conference on Computational Imaging Using Synthetic
  Apertures (CISA)}, pages 01--05. IEEE, 2024.

\bibitem{deshmane2012parallel}
Anagha Deshmane, Vikas Gulani, Mark~A Griswold, and Nicole Seiberlich.
\newblock Parallel {MR} imaging.
\newblock {\em Journal of Magnetic Resonance Imaging}, 36(1):55--72, 2012.

\bibitem{pruessmann2001advances}
Klaas~P Pruessmann, Markus Weiger, Peter B{\"o}rnert, and Peter Boesiger.
\newblock Advances in sensitivity encoding with arbitrary k-space trajectories.
\newblock {\em Magnetic Resonance in Medicine}, 46(4):638--651, 2001.

\bibitem{pruessmann2006encoding}
Klaas~P Pruessmann.
\newblock Encoding and reconstruction in parallel {MRI}.
\newblock {\em NMR in Biomedicine}, 19(3):288--299, 2006.

\bibitem{donoho2005stable}
David~L Donoho, Michael Elad, and Vladimir~N Temlyakov.
\newblock Stable recovery of sparse overcomplete representations in the
  presence of noise.
\newblock {\em IEEE Transactions on information theory}, 52(1):6--18, 2005.

\bibitem{lustig2007sparse}
Michael Lustig, David Donoho, and John~M Pauly.
\newblock Sparse {MRI}: The application of compressed sensing for rapid {MR}
  imaging.
\newblock {\em Magnetic Resonance in Medicine}, 58(6):1182--1195, 2007.

\bibitem{candes2008introduction}
Emmanuel~J Cand{\`e}s and Michael~B Wakin.
\newblock An introduction to compressive sampling.
\newblock {\em IEEE signal processing magazine}, 25(2):21--30, 2008.

\bibitem{plevritis1995spectral}
Sylvia~K Plevritis and Albert Macovski.
\newblock Spectral extrapolation of spatially bounded images.
\newblock {\em IEEE transactions on medical imaging}, 14(3):487--497, 1995.

\bibitem{samsonov2004pocsense}
Alexei~A Samsonov, Eugene~G Kholmovski, Dennis~L Parker, and Chris~R Johnson.
\newblock {POCSENSE}: {POCS}-based reconstruction for sensitivity encoded
  magnetic resonance imaging.
\newblock {\em Magnetic Resonance in Medicine}, 52(6):1397--1406, 2004.

\bibitem{peng2011reference}
Xi~Peng, Hui-Qian Du, Fan Lam, S~Derin Babacan, and Zhi-Pei Liang.
\newblock Reference-driven {MR} image reconstruction with sparsity and support
  constraints.
\newblock In {\em IEEE International Symposium on Biomedical Imaging}, pages
  89--92. IEEE, 2011.

\bibitem{dwork2024reducing}
Nicholas Dwork, Erin~K Englund, and Alex~J Barker.
\newblock Reducing the sampling burden of fourier sensing with a
  non-rectangular field-of-view.
\newblock {\em Applied Mathematics for Modern Challenges}, 4:1--16, 2025.

\bibitem{sodickson1997simultaneous}
Daniel~K Sodickson and Warren~J Manning.
\newblock Simultaneous acquisition of spatial harmonics ({SMASH}): fast imaging
  with radiofrequency coil arrays.
\newblock {\em Magnetic resonance in medicine}, 38(4):591--603, 1997.

\bibitem{griswold2002generalized}
Mark~A Griswold, Peter~M Jakob, Robin~M Heidemann, Mathias Nittka, Vladimir
  Jellus, Jianmin Wang, Berthold Kiefer, and Axel Haase.
\newblock Generalized autocalibrating partially parallel acquisitions
  ({GRAPPA}).
\newblock {\em Magnetic Resonance in Medicine}, 47(6):1202--1210, 2002.

\bibitem{haldar2020linear}
Justin~P Haldar and Kawin Setsompop.
\newblock Linear predictability in magnetic resonance imaging reconstruction:
  Leveraging shift-invariant fourier structure for faster and better imaging.
\newblock {\em IEEE Signal Processing Magazine}, 37(1):69--82, 2020.

\bibitem{mcmanus2025dependence}
Alex McManus, Stephen Becker, and Nicholas Dwork.
\newblock Dependence of parallel imaging with linear predictability on the
  undersampling direction.
\newblock {\em Journal of Electronic Imaging}, 34(2):023031--023031, 2025.

\bibitem{pruessmann1999sense}
Klaas~P Pruessmann, Markus Weiger, Markus~B Scheidegger, and Peter Boesiger.
\newblock {SENSE}: sensitivity encoding for fast {MRI}.
\newblock {\em Magnetic Resonance in Medicine}, 42(5):952--962, 1999.

\bibitem{pruessmann2001sensitivity}
Klaas~P Pruessmann, Markus Weiger, and Peter Boesiger.
\newblock Sensitivity encoded cardiac mri.
\newblock {\em Journal of cardiovascular magnetic resonance}, 3(1):1--9, 2001.

\bibitem{holme2019enlive}
H~Christian~M Holme, Sebastian Rosenzweig, Frank Ong, Robin~N Wilke, Michael
  Lustig, and Martin Uecker.
\newblock {ENLIVE}: an efficient nonlinear method for calibrationless and
  robust parallel imaging.
\newblock {\em Scientific reports}, 9(1):3034, 2019.

\bibitem{fessler2020optimization}
Jeffrey~A Fessler.
\newblock Optimization methods for magnetic resonance image reconstruction: Key
  models and optimization algorithms.
\newblock {\em IEEE signal processing magazine}, 37(1):33--40, 2020.

\bibitem{uecker2014espirit}
Martin Uecker, Peng Lai, Mark~J Murphy, Patrick Virtue, Michael Elad, John~M
  Pauly, Shreyas~S Vasanawala, and Michael Lustig.
\newblock {ESPIRiT}—an eigenvalue approach to autocalibrating parallel {MRI}:
  where {SENSE} meets {GRAPPA}.
\newblock {\em Magnetic resonance in medicine}, 71(3):990--1001, 2014.

\bibitem{lobos2023new}
Rodrigo~A Lobos, Chin-Cheng Chan, and Justin~P Haldar.
\newblock New theory and faster computations for subspace-based sensitivity map
  estimation in multichannel {MRI}.
\newblock {\em IEEE Transactions on Medical Imaging}, 43(1):286--296, 2023.

\bibitem{uecker2016bart}
Martin Uecker, Jonathan~I Tamir, Frank Ong, and Michael Lustig.
\newblock The {BART} toolbox for computational magnetic resonance imaging.
\newblock In {\em Proceedings of the International Society for Magnetic
  Resonance in Medicine}, volume~24, page~1, 2016.

\bibitem{lustig2010spirit}
Michael Lustig and John~M Pauly.
\newblock {SPIRiT}: iterative self-consistent parallel imaging reconstruction
  from arbitrary k-space.
\newblock {\em Magnetic resonance in medicine}, 64(2):457--471, 2010.

\bibitem{beatty2007method}
PJ~Beatty, AC~Brau, S~Chang, S~Joshi, CR~Michelich, E~Bayram, TE~Nelson,
  RJ~Herfkens, and JH~Brittain.
\newblock A method for autocalibrating {2D} accelerated volumetric parallel
  imaging with clinically practical reconstruction times.
\newblock In {\em Proceedings of the International Society for Magnetic
  Resonance in Medicine}, volume~15, page 1749, 2007.

\bibitem{majumdar2012choice}
Angshul Majumdar and Rabab~K Ward.
\newblock On the choice of compressed sensing priors and sparsifying transforms
  for mr image reconstruction: an experimental study.
\newblock {\em Signal Processing: Image Communication}, 27(9):1035--1048, 2012.

\bibitem{dutt1993fast}
Alok Dutt and Vladimir Rokhlin.
\newblock Fast fourier transforms for nonequispaced data.
\newblock {\em SIAM Journal on Scientific computing}, 14(6):1368--1393, 1993.

\bibitem{dutt1995fast}
Alok Dutt and Vladimir Rokhlin.
\newblock Fast fourier transforms for nonequispaced data, {II}.
\newblock {\em Applied and Computational Harmonic Analysis}, 2(1):85--100,
  1995.

\bibitem{dwork2023optimization}
Nicholas Dwork, Daniel O'Connor, Ethan~MI Johnson, Corey~A Baron, Jeremy~W
  Gordon, John~M Pauly, and Peder~EZ Larson.
\newblock Optimization in the space domain for density compensation with the
  nonuniform {FFT}.
\newblock {\em Magnetic resonance imaging}, 100:102--111, 2023.

\bibitem{bracewell1995two}
Ronald~N Bracewell.
\newblock {\em Two-dimensional imaging}.
\newblock Prentice-Hall, Inc., 1995.

\bibitem{ruderman1997origins}
Daniel~L Ruderman.
\newblock Origins of scaling in natural images.
\newblock {\em Vision research}, 37(23):3385--3398, 1997.

\bibitem{van1996modelling}
van~A Van~der Schaaf and JH~van van Hateren.
\newblock Modelling the power spectra of natural images: statistics and
  information.
\newblock {\em Vision research}, 36(17):2759--2770, 1996.

\bibitem{boyd2018introduction}
Stephen Boyd and Lieven Vandenberghe.
\newblock {\em Introduction to applied linear algebra: vectors, matrices, and
  least squares}.
\newblock Cambridge university press, 2018.

\bibitem{beck2009fast}
Amir Beck and Marc Teboulle.
\newblock A fast iterative shrinkage-thresholding algorithm for linear inverse
  problems.
\newblock {\em SIAM journal on imaging sciences}, 2(1):183--202, 2009.

\bibitem{o2018fraction}
Daniel O’Connor, Victoria Yu, Dan Nguyen, Dan Ruan, and Ke~Sheng.
\newblock Fraction-variant beam orientation optimization for non-coplanar
  {IMRT}.
\newblock {\em Physics in Medicine \& Biology}, 63(4):045015, 2018.

\bibitem{combettes2005signal}
Patrick~L Combettes and Val{\'e}rie~R Wajs.
\newblock Signal recovery by proximal forward-backward splitting.
\newblock {\em Multiscale modeling \& simulation}, 4(4):1168--1200, 2005.

\bibitem{ong2018mridata}
F~Ong, S~Amin, S~Vasanawala, and M~Lustig.
\newblock Mridata.org: An open archive for sharing {MRI} raw data.
\newblock In {\em Proceedings of the International Society for Magnetic
  Resonance in Medicine}, volume~26, 2018.
\newblock \url{www.mridata.org}.

\bibitem{dwork2024accelerated}
Nicholas Dwork, Jeremy~W Gordon, and Erin~K Englund.
\newblock Accelerated parallel magnetic resonance imaging with compressed
  sensing using structured sparsity.
\newblock {\em Journal of Medical Imaging}, 11(3):033504--033504, 2024.

\bibitem{dwork2021fast}
Nicholas Dwork, Corey~A Baron, Ethan~MI Johnson, Daniel O'Connor, John~M Pauly,
  and Peder~EZ Larson.
\newblock Fast variable density poisson-disc sample generation with directional
  variation for compressed sensing in {MRI}.
\newblock {\em Magnetic resonance imaging}, 77:186--193, 2021.

\bibitem{adcock2017breaking}
Ben Adcock, Anders~C Hansen, Clarice Poon, and Bogdan Roman.
\newblock Breaking the coherence barrier: A new theory for compressed sensing.
\newblock In {\em Forum of mathematics, sigma}, volume~5, page~e4. Cambridge
  University Press, 2017.

\bibitem{dwork2021utilizing}
Nicholas Dwork, Daniel O’Connor, Corey~A Baron, Ethan~MI Johnson, Adam~B
  Kerr, John~M Pauly, and Peder~EZ Larson.
\newblock Utilizing the wavelet transform’s structure in compressed sensing.
\newblock {\em Signal, image and video processing}, 15:1407--1414, 2021.

\bibitem{dwork2022utilizing}
Nicholas Dwork and Peder~EZ Larson.
\newblock Utilizing the structure of a redundant dictionary comprised of
  wavelets and curvelets with compressed sensing.
\newblock {\em Journal of Electronic Imaging}, 31(6):063043--063043, 2022.

\bibitem{wang2004image}
Zhou Wang, Alan~C Bovik, Hamid~R Sheikh, and Eero~P Simoncelli.
\newblock Image quality assessment: from error visibility to structural
  similarity.
\newblock {\em IEEE transactions on image processing}, 13(4):600--612, 2004.

\bibitem{ong2018low}
Frank Ong.
\newblock {\em Low dimensional methods for high dimensional magnetic resonance
  imaging}.
\newblock PhD thesis, University of California, Berkeley, 2018.

\bibitem{chambolle2011first}
Antonin Chambolle and Thomas Pock.
\newblock A first-order primal-dual algorithm for convex problems with
  applications to imaging.
\newblock {\em Journal of mathematical imaging and vision}, 40(1):120--145,
  2011.

\bibitem{malitsky2018first}
Yura Malitsky and Thomas Pock.
\newblock A first-order primal-dual algorithm with linesearch.
\newblock {\em SIAM Journal on Optimization}, 28(1):411--432, 2018.

\bibitem{condat2013primal}
Laurent Condat.
\newblock A primal--dual splitting method for convex optimization involving
  lipschitzian, proximable and linear composite terms.
\newblock {\em Journal of optimization theory and applications},
  158(2):460--479, 2013.

\bibitem{yan2016primal}
Ming Yan.
\newblock A primal-dual three-operator splitting scheme.
\newblock Technical report, Technical Report, 2016.

\end{thebibliography}

\end{document}